\documentclass[letter,dvipsnames]{aa}  

\usepackage{graphicx}
\usepackage{txfonts}
\usepackage{url}
\usepackage{hyperref}
\hypersetup{
    colorlinks=true,
    linkcolor=blue,
    urlcolor=magenta,
}
\usepackage{listings}
\definecolor{dkgreen}{rgb}{0,0.6,0}
\definecolor{gray}{rgb}{0.5,0.5,0.5}
\definecolor{mauve}{rgb}{0.58,0,0.82}

\usepackage{orcidlink}

\begin{document}

   \title{Spiral-like features in the disc revealed by Gaia DR3 radial actions}


   \author{P. A. Palicio \inst{1}\orcidlink{0000-0002-7432-8709} \and A. Recio-Blanco \inst{1}\orcidlink{0000-0002-6550-7377}
         \and E. Poggio \inst{1,2}\orcidlink{0000-0003-3793-8505} \and T. Antoja  \inst{3,4,5}\orcidlink{0000-0003-2595-5148}\and
          P. J. McMillan \inst{6}\orcidlink{0000-0002-8861-2620} \and E. Spitoni \inst{1}\orcidlink{0000-0001-9715-5727}
          }

   \institute{Université Côte d’Azur, Observatoire de la Côte d’Azur, CNRS, Laboratoire Lagrange, France\\
              \email{pedro.alonso-palicio@oca.eu}
              \and
            {Osservatorio Astrofisico di Torino, Istituto Nazionale di Astrofisica (INAF), I-10025 Pino Torinese, Italy}
            \and
             {Departament de Física Qu\`antica i Astrof\'isica (FQA), Universitat de Barcelona (UB),  c. Mart\'i i Franqu\`es, 1, 08028 Barcelona, Spain}
             \and {Institut de Ci\`encies del Cosmos (ICCUB), Universitat de Barcelona (UB), c. Mart\'i i Franqu\`es, 1, 08028 Barcelona, Spain} 
             \and {Institut d'Estudis Espacials de Catalunya (IEEC), c. Gran Capit\`a, 2-4, 08034 Barcelona, Spain}
            \and
            {Lund Observatory, Department of Astronomy and Theoretical Physics, Lund University, Box 43, SE-22100, Lund, Sweden}
             }

    \date{Received XXXX; accepted YYY}

 
  \abstract
   {The so-called action variables are specific functions of the positions and velocities that remain constant along the stellar orbit. The astrometry provided by \textit{Gaia} Early Data Release 3 (EDR3), combined with the velocities inferred from the RVS (Radial Velocity Spectrograph) spectra of \textit{Gaia} DR3, allows the estimation of these actions for the largest volume of stars to date.}
   {We aim to explore these actions to find structures in the Galactic disc.}
   {We compute the actions and the orbital parameters of the \textit{Gaia} DR3 stars assuming an axisymmetric model for the Milky Way. Using \textit{Gaia} DR3 photometric data, we select a subset of giant stars with better astrometry as a control sample.}
   {The maps of the percentiles of the radial action $J_R$ reveal arc-like segments. We find a high $J_R$ region centered at $R\approx10.5$~kpc of $1$~kpc width, as well as three arc-shape regions dominated by circular orbits at inner radii. We also identify the spiral arms in the overdensities of the giant population.}
   {For Galactic coordinates (X, Y, Z), we find good agreement with the literature in the innermost region for the Scutum-Sagittarius spiral arms. At larger radii, the low $J_R$ structure tracks the Local arm at negative $X$, while for the Perseus arm the agreement is restricted to the $X<2$~kpc region, with a displacement with respect to the literature at more negative longitudes. We detect a high $J_R$ area at a Galactocentric radii of $\sim 10.5$~kpc, consistent with some estimations of the Outer Lindblad Resonance location. We conclude that the pattern in the dynamics of the old stars is consistent in several places with spatial distribution of the spiral arms traced by young populations, with small potential contributions from the moving groups.}


   \authorrunning{Palicio et al.}

   \maketitle
%

\section{Introduction}
\par The \textit{Gaia} satellite \citep{TheGaiaMission16, de_Bruijne2012, Gaia2018, GaiaDR3} constitutes the most advanced astrometric mission to date. After its launch in 2013, it has been providing positions, parallaxes, proper motions and line-of-sight velocities for an increasing number of sources in subsequent data releases \citep{Katz04, Cropper18, Katz19, Katz22}. This exquisite astrometry has improved our understanding of Galactic structures already known, like the spiral arms and the warp \citep{Antoja16, Poggio18, Poggio21, Chrobakova22, PVPCCoMW}, and revealed a complex formation for the Milky Way in strong interaction with other galaxies \citep{Belokurov18, Myeong18, Myeong19, Helmi2018,Koppelman_et_al19, Helmi_20}. In this context, many asymmetries in different parameter spaces have been interpreted as a consequence of this scenario, including velocities \citep{Antoja17}, distribution of proper motions \citep{Palicio20}, ridges in projected velocities \citep{Ramos18, Fragkoudi19, Khoperskov21, PVPCCoMW, PVPDimmel22, McMillan22}, distribution of actions \citep{Hunt19, Sellwood19, BlandHawthorn19, Trick19, Trick21, Trick22} and distribution of metallicity \citep{Poggio22}. In this work, we report the structures in the Galactic plane revealed by the distribution of the radial action $J_R$ computed with the \textit{Gaia} DR3 astrometry and line-of-sight velocities \citep{GaiaDR3}.
\par This Letter is organised as follows: in Section \ref{Sec_Data} we explain the selection criteria applied to the \textit{Gaia} data of our sample. In Section \ref{Sec_OrbParam} we describe the model and the performance adopted for computing the orbital parameters and actions from the input data. Results are shown and discussed in Sections \ref{Sec_Results} and \ref{Sec_Discussion}, respectively. The conclusions can be found in Section \ref{Sec_conclusion}. Finally, in the Appendices \ref{Appendix:Query} and \ref{AppendixStackel} we specify the data query performed on the \textit{Gaia} archive\footnote{\url{https://gea.esac.esa.int/archive/}} and the detailed procedure for the estimation of the actions and orbital parameters, respectively. In Appendix \ref{Appendix:Giants} we reproduce our analysis with a subsample of giants stars selected by photometry.
%
%
%
%
%
%
%
\begin{figure*}[!h]
\centering
\includegraphics[width=0.70\textwidth]{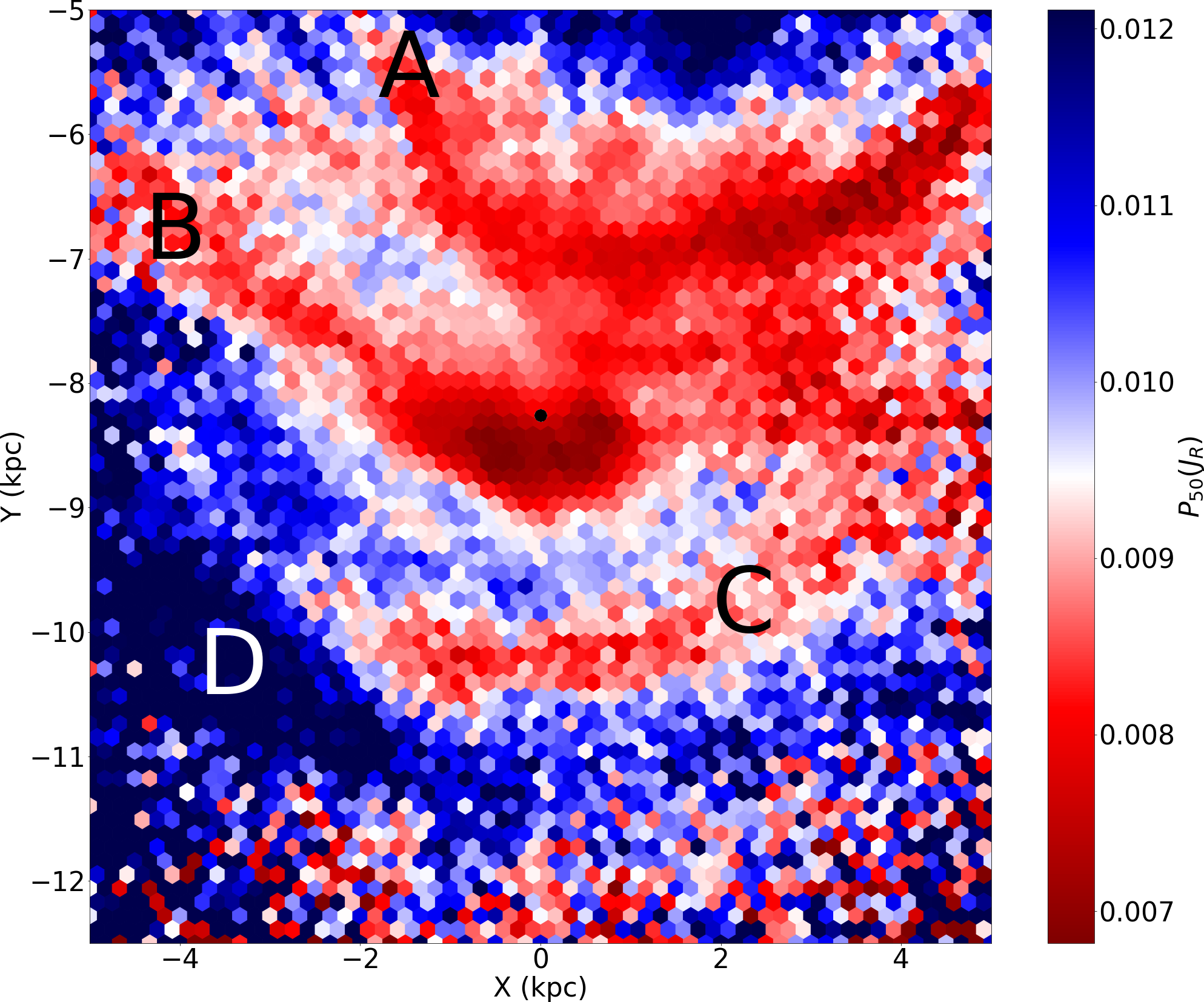}
\caption{Distribution of the median (equivalent to the $P_{50}$ percentile) of $J_R$  on the Galactic Plane ($|Z_{max}|<0.5$~kpc). The solid black circle denotes the solar position. The features discussed in the text are labelled from A to D.}
\label{Fig_medianJR}
\end{figure*}
\begin{figure*}[h]
\includegraphics[width=0.98\textwidth]{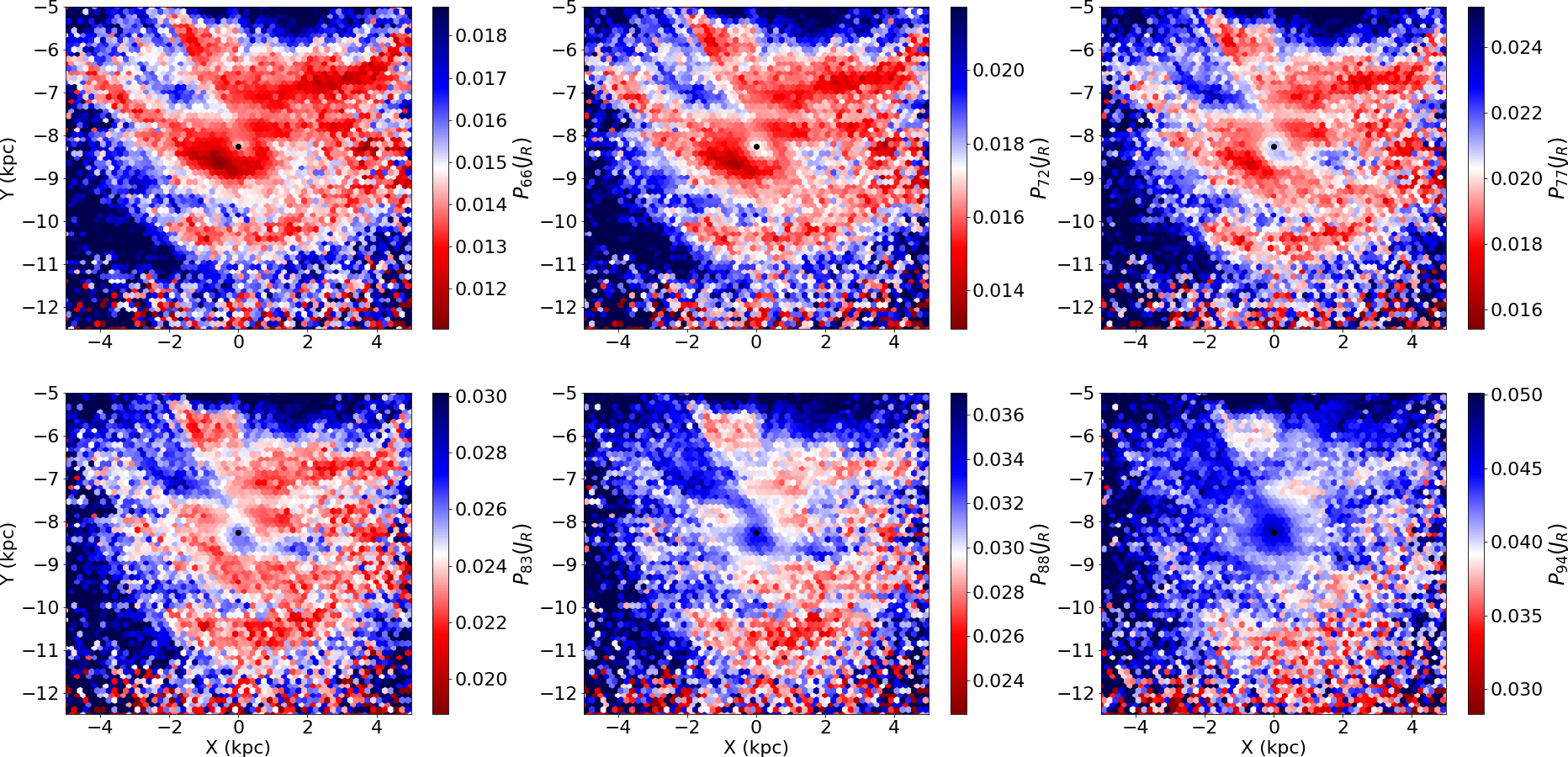}
\caption{Distribution of $J_R$ percentiles on the Galactic Plane ($|Z_{max}|<0.5$~kpc). The $x$ percentile is denoted by $P_x$. The solid black circle denotes the solar position.}
\label{Fig_percentJR}
\end{figure*}
%
%
%
\section{Gaia data and selection criteria}
\label{Sec_Data}
\par We make use of all the \textit{Gaia} DR3 stars with full astrometric information available (parallaxes, positions, proper motions and line of sight velocities) and select those with non null geometric distance estimation \citep{Bailer-Jones21}. The corresponding \texttt{ADQL} query can be found in Appendix \ref{Appendix:Query}. This sample totals $33,653,049$ million sources, in which we select those with good kinematic measurements by imposing a maximum line-of-sight velocity error of $5$~km/s and a relative error in proper motion lower than $15\%$. For the heliocentric distance, we impose a maximum relative error of 20\%. Since we focus our study on the disc, we exclude those stars whose maximum distance from the Galactic plane is larger than $500$~pc (see Section \ref{Sec_OrbParam}). The resulting sample size is $\sim12.4$ million sources. We correct the line-of-sight velocities and proper motions assuming $(U_\odot, V_\odot, W_\odot)=(9.5, 250.7, 8.56)$~km/s for the solar motion \citep{Gravity20, Reid20} and $R_{\odot}=8.249$~kpc for the Galactocentric distance of the Sun \citep[][]{Gravity20}.
\par In order to propagate the errors, we consider the correlations between the astrometric parameters. Motivated by the approach of \citet{PVPCCoMW} and \citet{Kordopatis22}, we model the distribution of the errors of the geometric distances with a broken Gaussian distribution parameterised by the input confidence intervals (i.e, $r_{lo}$ and $r_{hi}$ in \citealt{Bailer-Jones21}).
\section{Orbital parameters and actions}
\label{Sec_OrbParam}
\par We model the forces of the Milky Way with a rescaled version of the potential of \citet{McMillan17} such that the circular velocity at $R_\odot=8.249$~kpc is $V_\odot=238.5$~km/s, consistent with our assumed Solar motion and the velocity of the Sun with respect to the Local Standard of Rest taken from \citet{SchBinDeh2010}. This potential is fully axisymmetric and models the contribution of the halo, bulge, thin and thick stellar discs as well as the H\,\textsc{I} and H\,\textsc{II} gas discs. We estimate the orbital parameters (apocenter $r_{apo}$, pericenter $r_{peri}$, maximum orbital distance to the galactic plane $Z_{max}$) and the non-trivial actions $J_R$ and $J_Z$ by using own implementation of the Stäckel-Fudge approximation \citep{Binney12, Sanders16, Mackereth18}. We refer to Appendix \ref{AppendixStackel} for a detailed description of this procedure. Apart from these parameters, the vertical component of the angular momentum $L_z$ and the total energy $E$ are obtained as output. The actions $J_R$, $J_Z$ and the angular momentum $L_z$ presented in this work are expressed in units of $L_\odot=R_\odot V_\odot$. The resulting data table will be published online.
\section{Results}
\label{Sec_Results}
\par In this Section we explore the map of the distribution of the radial action $J_R$ in the Galactic Plane. Figure \ref{Fig_medianJR} shows the spatial distribution of the median $J_R$, while each panel in Figure \ref{Fig_percentJR} refers to other percentiles to illustrate the variation of the distribution of $J_R$ across the Galactic plane. Due to the variations of the observed trends as a function of the considered percentile, the colorbar is tuned to enhance the contrast between the high and low $J_R$ regions in each panel. We identify three main structures in the low $J_R$ regions for the first four percentiles shown in the figure (labelled as A, B and C in Fig. \ref{Fig_medianJR}), while for the $94$-th percentile they are highly distorted. We observe an additional feature (labelled as D) in the outer part of the disc ($10$~kpc $\lesssim R\lesssim 11$~kpc) characterised by high $J_R$ values.
\par The innermost structure, labelled as A, extends from $R\approx 6.0~$kpc at ($X$, $Y$)$\approx$($-2$, $-5.5$)~kpc to $R\sim7$~kpc at the solar azimuth ($X=0$~kpc direction), while for $X<0$~kpc it shows an almost constant radii of $R\approx7-7.2$~kpc. This results in a longitudinally asymmetric arc-shape structure of variable pitch angle.
\par Structure B also shows significant variations with longitude\footnote{We denote the Galactic longitude and latitude with ($\ell$, $b$), respectively, where $\ell$ increases counter-clockwise from the Sun-Galactic centre direction.}: for $X<0$ we observe a well defined low $J_R$ area that extends from ($X$, $Y$)$\approx$($-4$, $-6.5$)~kpc to ($0$, $8.5$)~kpc, embedding the solar neighbourhood. However, its prolongation at negative $X$ is highly distorted, resulting in a wide area of low $J_R$ between Structure A and the $\ell=-90^\circ$ direction.
\par In contrast to the previous features, Structure C is sharply defined at negative $X$, where it extends from ($X$, $Y$)$\approx$($4$, $-9$)~kpc to ($-2$, $-10$)~kpc, although it is possible to discern a tail of relatively low $J_R$ at $X<-2$~kpc. At positive $X$, this structure is connected with one of the extensions of the feature B, located in the large low $J_R$ area found between A and B ($X>0$~kpc and $-8\lesssim Y\lesssim-7$~kpc), and creating a gap of high $J_R$ with B. As can be seen in Fig. \ref{Fig_percentJR}, this feature extends towards outer radii for percentiles larger than $P_{83}$, and constitutes the only low $J_R$ structure at large percentile ($P_{94}$).
\par The outermost feature (D) is a high $J_R$ region with an arc-shape of almost constant radii of $10.5$~kpc and $\sim 1.0$~kpc width. It remains almost unchanged for percentiles lower than $P_{77}$ and becomes blurred for higher values.
\par Apart from the main features, it is worthwhile mentioning the bifurcation in Structure A at ($X\lesssim -2$~kpc, $Y\approx-6$~kpc) towards positive $X$, although it gets distorted in the maps for the large percentiles ($P_{66}$ and above). Finally, we can discern a subtle arc-shape structure between A and B with very low median radial action ($P_{50}<0.008$) from ($X$, $Y$)$\approx$($0$, $-7.7$) to ($2$, $-7.7$).
\begin{figure}[h]
\centering
\includegraphics[width=0.48\textwidth]{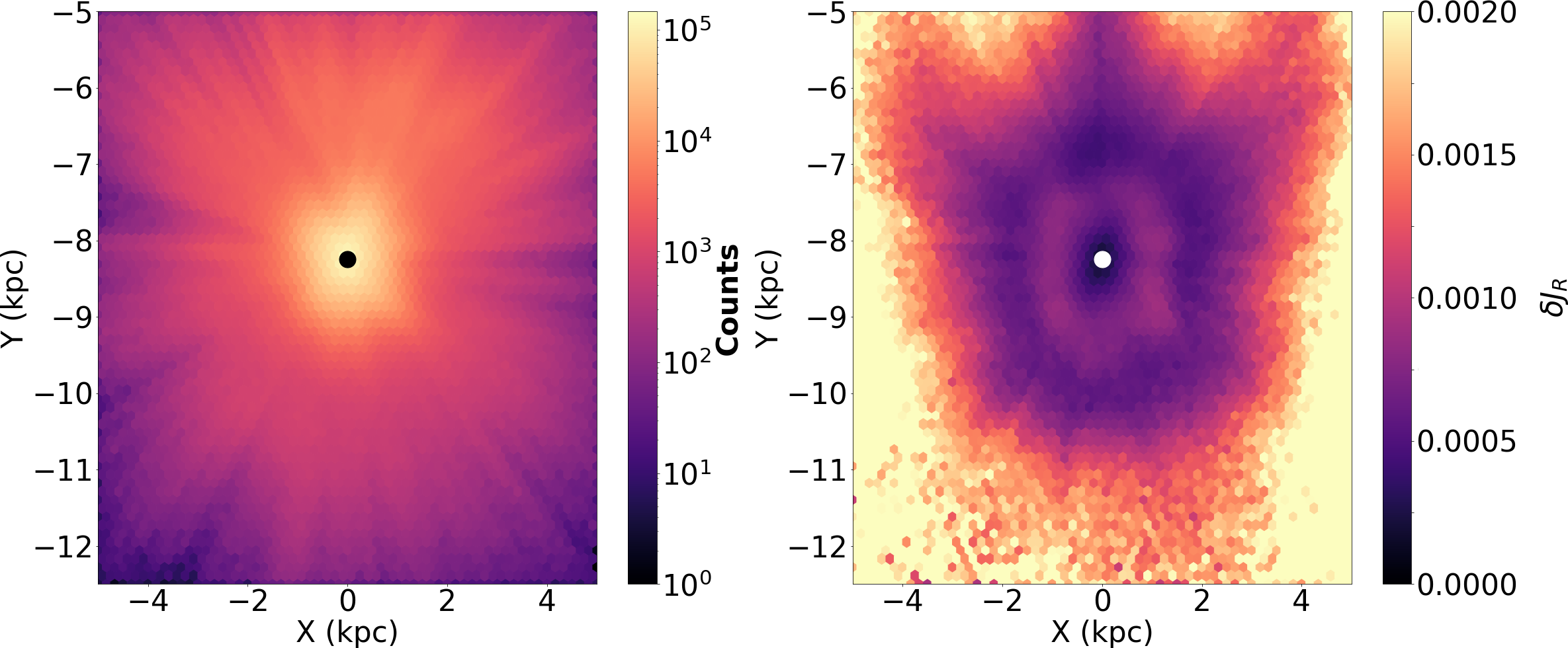}
\caption{Left panel: Density map of the selected sample in the Cartesian plane $(X, Y)$. Right panel: distribution of median errors in $J_R$ on the Galactic plane.}
\label{Fig_densities}
\end{figure}
\begin{figure*}[h]
\includegraphics[width=0.99\textwidth]{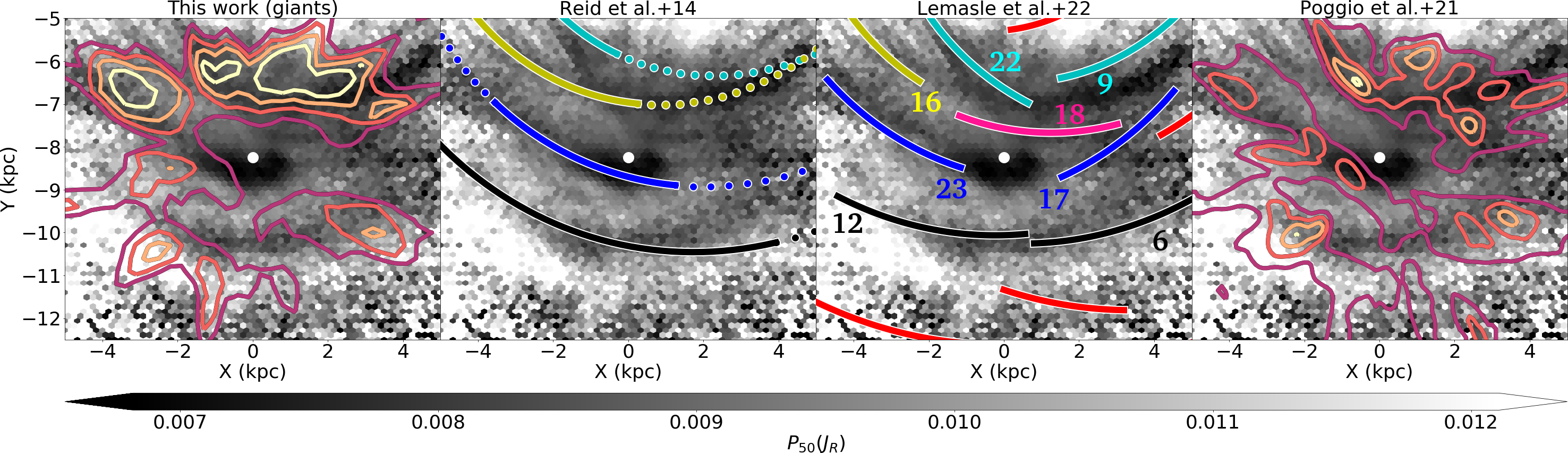}
\caption{Comparison of the maps of $P_{50}(J_R)$ with the spiral arms reported in literature. First panel: contour lines enclose the overdensities found in the subsample of giants. Second panel: solid lines represent the Scutum (cyan), Sagittarius (yellow), Local (blue) and Perseus (black) spiral arms of \citet{Reid14}, while dotted lines correspond to their extrapolation in azimuth. Third panel: solid lines represent the segments of spiral arms of \citet{Lemasle22}, in which their same naming convention is used, while the colorcode results from a visual comparison with these of \citet{Reid14}. The additional structures are indicated by red and pink lines for description convenience. Fourth panel: contour lines illustrate the overdensities reported by \citet{Poggio21}. Background image: reproduction of Fig. \ref{Fig_medianJR} using a grey color-scale to increase the contrast between the coloured lines and the background map. Solid white circle denotes the solar position.}
\label{Fig_fourpanels}
\end{figure*}
\par In order to check if the features described above are a consequence of the distribution of stars, we represent in Figure \ref{Fig_densities} the density map of the selected sample. As it can be seen in the left panel, the density map cannot explain all the structures identified in Figures \ref{Fig_medianJR} and \ref{Fig_percentJR}. The density map peaks at the solar position and decreases with the heliocentric distance as fainter stars are excluded, showing no correspondence with the arc-shaped structures in the $J_R$ distribution.
\par We verify the significance of the features in $J_R$ with the observational errors by evaluating the map of the median error of the radial action, $\delta J_R$, estimated from 25 realisations of the input data (right panel in Fig. \ref{Fig_densities}). Although it is possible to distinguish some selection effects in an annular region centred at the Sun, the structure associated with them does not correspond to that reported in Figures \ref{Fig_medianJR} and \ref{Fig_percentJR}. Furthermore, in the vast majority of the plane, the errors of $J_R$ are at least $3.5$ times smaller than the median $J_R$, supporting the robustness of the features found in the percentile distributions.
\section{Discussion}
\label{Sec_Discussion}
\par In this Section, we discuss three possible scenarios to explain the observed features in $J_R$. 
\subsection{Spiral Arms}
\label{Subsec_spiral}
The spatial distribution and shape of the structures reported above suggest a connection with the spiral arms. To explore this hypothesis, we compare these structures with the fit of the spiral arms inferred from the kinematics of one hundred masers \citep{Reid14}, from the distribution of Cepheids \citep{Lemasle22} and from the distribution of Gaia EDR3 Upper Main Sequence stars \citep[UMS stars,][]{Poggio21}, which considers the same astrometric measurements (but for a different sample) as this work. We complement these references with the overdensity map of our subsample of giant stars (see Appendix \ref{Appendix:Giants}). Following the procedure described in \citet{Poggio21}, we compute the local (average) density using an Epanechnikov kernel \citep{Epanechnikov69} of bandwidth 0.3~kpc (2.0~kpc). This kernel assumes that the contribution of a star to the density at a reference point is weighted by a term $\propto \textnormal{max}\lbrace 1-x^2/h^2, 0\rbrace$, where $h$ is the bandwidth and $x$ is the separation between the star and the reference point. For sake of visualisation, the references of spiral arms described above are shown in individual panels in Figure \ref{Fig_fourpanels}.
\par First panel in Fig. \ref{Fig_fourpanels} illustrates the overdensities in the distribution of our sample of giants. We find a correspondence between the overdensities in this sample and these reported by \citet[][]{Poggio21} for the younger UMS population (fourth panel); though discrepancies are observed at $(X,Y)\approx(-1, -9)$~kpc and ($-2$,$-6.5$)~kpc. The presence of the spiral arms traced by an old population has been recently proposed by \citet{Lin2022}, who identify the Local Arm in a sample of $~87,000$ \textit{Gaia} EDR3-2MASS \citep{GaiaEDR3, TWOMASSpaper} Red Clump stars (RC), and could be related to the metallicity asymmetry in Sample B and C in \citet[][see their figure 1]{Poggio22}.
\par In general terms, we find a good agreement between the low $J_R$ areas and the spiral arms, especially in the innermost regions, where the distribution of giant stars (first panel) reveals an overdensity consistent with Structure A. Furthermore, the bifurcation observed in A can be explained by the segments 16 and 22 of \citet{Lemasle22}, likely to be part of Sagittarius and Scutum respectively. On the contrary, we find a shift of $\sim0.5$~kpc between Structure A and the location of the segment 9, where the extrapolation of \citet{Reid14} perfectly fits the lowest $J_R$ region of A (dotted lines). Compared to \citet{Poggio21}, we can identify most of Structure A in the innermost overdensity of UMS stars, though no bifurcation is observe at $X\approx-2$~kpc. The extension of this overdensity, however, is compatible with the area of low $J_R$ that connects the structures A and B in Fig. \ref{Fig_medianJR}. 
\par As mentioned in Section \ref{Sec_Results}, we find a subtle arc-shape structure between A and B close to the solar neighbourhood. This feature has no counterpart in the spiral arms of \citet{Reid14}, \citet{Poggio21} or in our distribution of giant stars, but it is located at the same position as the segment 18 (pink line) of \citet{Lemasle22}, being a potential continuation of the Sagittarius arm. As the percentile increases (Fig. \ref{Fig_percentJR}), this small low $J_R$ area becomes more evident (a gap with A emerges) and consistent with the segment 18 and its extension towards positive $X$.
\par The part of Structure B located at negative $X$ is compatible with the fit of the Local spiral arm of \citet{Reid14}, the segment 23 of \citet{Lemasle22} and the overdense regions found in the UMS and giant population. On the contrary, at positive $X$ only the Local arm of \citet{Reid14} might provide a good explanation for Structure B, but only if a shift of $\sim0.5$~kpc is considered. It is worthwhile mentioning the significant differences among the references for that part of the Local arm: assuming the segment 17 is part of the Local arm, it implies a pitch angle of opposite sign compared to that in \citet{Poggio21}, while according to \citet{Reid14} the Local arm is more tangential. This variety of observations suggests a complex definition of the extension and limits of the Local spiral arm despite its proximity to the Sun. 
\par The major discrepancy is found in the solar neighbourhood: according to our maps, the Sun is embedded in the intersection of the Local and the Sagittarius spiral arm, while the predictions of all three spiral arms maps report a solar location in the inner boundary of the Local Arm. 
\par The $J_R$ maps suggest a connection between the Perseus and the Local Arm (Structures C and B, respectively). However, the spatial geometry of the spiral arms from UMS stars \citep{Poggio21}, Red Clump stars \citep{Lin2022} and the giant sample does not coincide with the observed features in $J_R$ in this region. 
\par The comparison of Structure C reveals a good agreement with the Perseus spiral arm of \citet{Reid14} for $X\lesssim 0$~kpc and the segment 12 of \citet{Lemasle22} within $|X|\lesssim 1$~kpc. However, at positive $X$, Structure C exhibit a different pitch angle compared to both \citet{Reid14} and \citet{Lemasle22}.
\par It is worth mentioning that the spiral structure of the Milky Way might be different depending on the considered stellar population. Here, the contrast in $J_R$ is observed mainly in the giant old population (see Appendix \ref{Appendix:Giants} for the specific analysis of the giant stars), even though the stars in the spiral arms tend to be young and, through the age-velocity dispersion relation, show lower values of $J_R$. For instance, the referred spiral arms have been traced by selecting masers, Cepheids and Upper Main Sequence stars; that is, the young population. Thus, the dynamics of the old stars seem to be in agreement with the spatial distribution of the young population in some regions, but present some discrepancies in others. Such discrepancies can be either due to the fact that the geometry of the spiral arms might be different for different stellar populations, or that the dynamical nature of the spiral arms somehow leads to the observed features. 
%
%
%
%
%
%
\subsection{Moving groups}
\label{Subsec_moving_groups}
\par
\begin{figure}[h]
\includegraphics[width=0.5\textwidth]{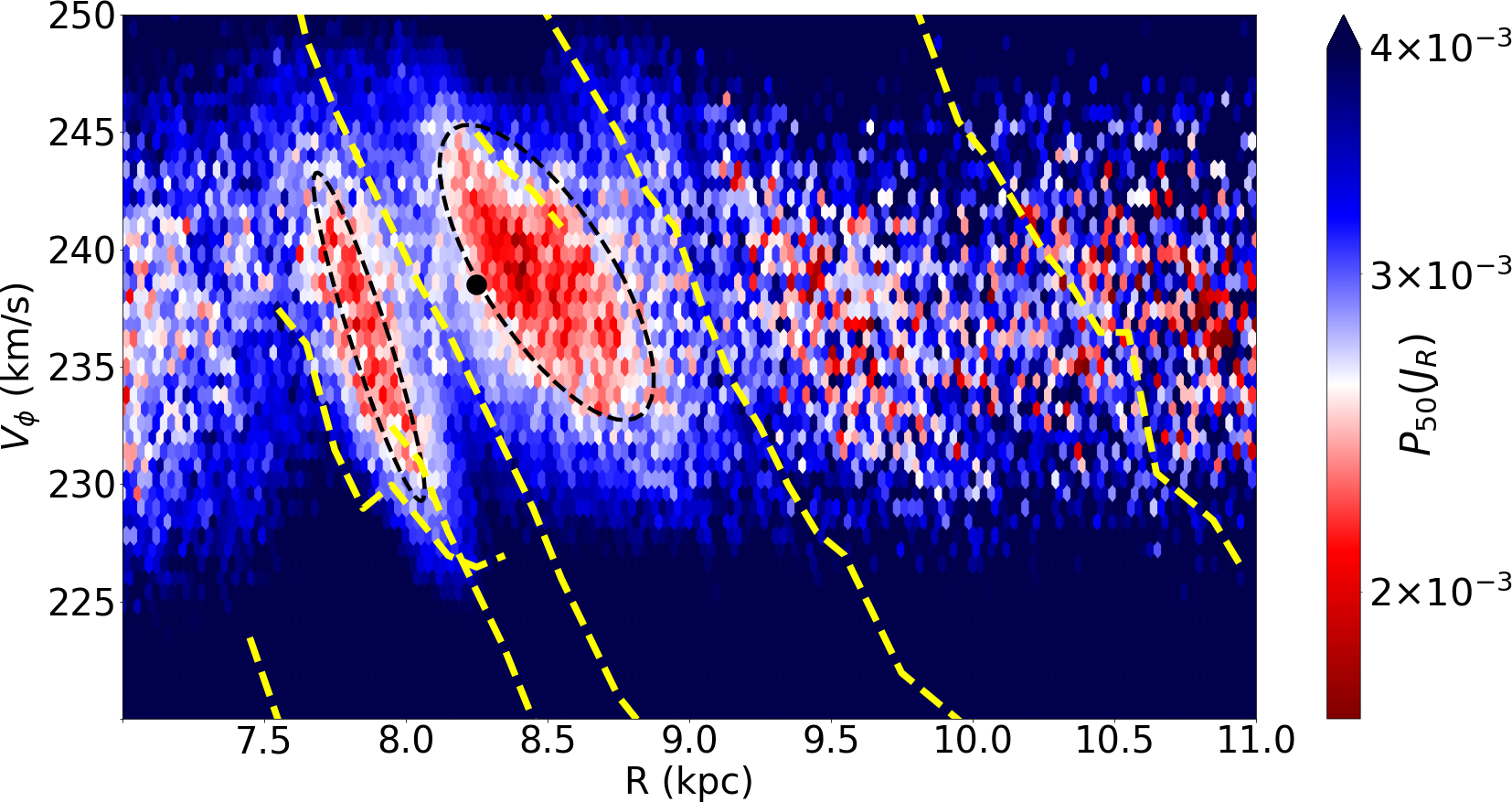}
\caption{Azimuthal velocity $V_{\phi}$ vs. $R$ diagram colorcoded with the median $J_R$. The colorbar has been intentionally set in logarithmic scale to cover a wide range of values in $J_R$. The moving groups (dashed yellow lines) are displayed from the bottom left to the upper right corner as follows: Hercules, Dehnen98-6, Horn-Dehnen98, Hyades, Coma Berenices, Sirius and Arch1-Hat. Black ellipses enclose the two selected areas (see the text) while the Sun is denoted by the solid black circle.}
\label{Fig_moving_groups}
\end{figure}

 We also explore the possible origin of the reported structures in the moving groups. As \citet{Ramos18} show, it is possible to identify the moving groups as stripes in the azimuthal velocity $V_{\phi}$ vs. $R$ diagram. Figure \ref{Fig_moving_groups} represents the distribution of the median $J_R$ in the ($R$,$V_{\phi}$) plane, including some of the moving groups reported by \citet{Ramos18} as reference (yellow dashed lines). For sake of visualisation, we focus on the range $220<V_{\phi}<250$~km/s and use a logarithmic colorscale for median($J_R$) to enhance the features. As expected, the values of $J_R$ tend to increase as $V_{\phi}$ differs from the rotation curve. As Figure \ref{Fig_moving_groups} shows, the Dehnen98-6, Hyades and Sirius moving groups are predominantly located in areas of relatively high $J_R$ in the ($R$,$V_{\phi}$) plane, in contrast to the low $J_R$ values that characterise the features described in Section \ref{Sec_Results}. On the contrary, Coma Berenices lies close to a transition from low to high $J_R$. The Hercules and most of the Horn-Dehnen98 moving groups lie in the region of high $J_R$ (blue saturated region) and we do not see any clear correspondence for the Arch1-Hat moving group at this point. In any case these groups could be related to the structures of $J_R$ in the $R$-$V_{\phi}$ projection that extend to higher $J_R$ (not seen in our figures due to the colour range). 
\par Apart from the ridges, we can identify two interesting areas of low $J_R$: one located between the Dehnen98-6 and the Hyades moving groups, as a prolongation of Horn-Dehnen98 at inner radii; and another more extended low $J_R$ area close to Coma Berenices. In order to evaluate the contribution of this potential members of moving groups, we exclude the stars within these regions (black dashed ellipses in Fig. \ref{Fig_moving_groups}). We have verified the exclusion of the stars close to Coma Berenices raises the median values of $J_R$ at $\sim$(2, -8)~kpc, improving the separation between the A and B structures at positive $X$. The exclusion of the other selection, however, leads to an annular distortion at $R\sim7.7$~kpc that increases the gap between the A and B structures, especially at negative $X$. This distortion, however, is more likely to be an artefact caused by the exclusion of a significant number of sources within $7.6 \lesssim R\lesssim 8.0$~kpc rather than a true contribution of the Horn-Dehnen moving group.
\par Based on our tests, the features observed in the $V_{\phi}$ vs. $R$ plane are not as clear as those found in the maps of $J_R$, although an apparent relation between the high $J_R$ values and the position of some ridges can be inferred. A deeper analysis of this relation is needed to evaluate the contribution of the moving groups to the features in $J_R(X,Y)$ and, potentially, its connection with the spiral arms. That analysis is beyond the scope of this Letter and will be explored in a future work.
%
%
%
%
%
%
%
%
%
\subsection{Galactic bar}
\label{Subsec_bar}
\par Apart from the spiral arms, the location and shape of the high $J_R$ region at $R\sim 10.5$~kpc is consistent with some values reported for the Outer Lindblad Resonance \citep[OLR;][]{Liu12, Portail17, PerezVillegas17}. In order to evaluate this possible connection we verify if the OLR corresponds to a region of increasing radial action. Under the epicyclic approximation \citep[see][]{Binney2008}, the Galactocentric distance $R(t)$ of a star trapped by a resonance varies with time as:
\begin{equation}
\label{Eq_R_of_t}
R(t)=R_g - C_2\cos{(2\Delta\Omega t)}\ \ \ \ (\Delta\Omega\equiv\Omega-\Omega_p)
\end{equation}
where the factor 2 in the cosine comes from the assumption of a dipolar disturbance of the potential ($m=2$), $R_g$ is the guiding radius, $\Omega$ is the circular frequency at $R_g$, $\Omega_p$ is the pattern speed and $C_2$ is a constant that depends on the bar potential $\Phi_b(R, \phi,t)$ as:
\begin{equation}
\label{Eq_C_2}
C_2 = \frac{1}{\kappa^2-4\Delta \Omega^2} \times \left( \left. \frac{d\Phi_b}{dR} + \frac{2\Omega \Phi_b}{R\Delta \Omega} \right) \right|_{Rg}  
\end{equation}
with $\kappa$ the epicyclic frequency at $R=R_g$.
Differentiating Eq. \ref{Eq_R_of_t} with respect to the time and substituting in the integral for $J_R$ (see Eq. \ref{Eq_Ju_Jv}) we have
\begin{equation}
\label{Eq_JR_epicyclic_definitive}
J_R = \frac{2 \Delta \Omega C_2}{\pi} \int_{R_g-C_2}^{R_g+C_2} \sin{(2\Delta\Omega t)} dR =  C_2^2\cdot \Delta\Omega 
\end{equation}
where the upper and lower limits correspond to the cases in which $\cos{(\Delta\Omega t)}=-1$ (apocenter) and $+1$ (pericenter), respectively. Thus, Eq. \ref{Eq_JR_epicyclic_definitive} diverges in the Outer Lindblad Resonance ($\kappa +2\Delta \Omega =0$). Although Eq. \ref{Eq_R_of_t} assumes a small deviation in azimuth with respect to the circular orbit defined by the guiding radius, which is not true in the resonance regime, it is enough to demonstrate the radial action increases towards the resonances \citep{Chiba2021}. A more detailed analysis, like that described for the Corotation in Section 3.3b of \citet{Binney2008} would predict a large but finite action. However, the calculus of this more general case is not straightforward \citep{Goldreich81}.
\par According to \citet{Sellwood2002} and \citet{Sellwood10}, not only the OLR but also the Inner Lindblad Resonance (ILR) should be characterised by an increment in $J_R$ due to the outward flow of angular momentum \citep{LyndenBell1972}. Assuming a pattern speed for the bar between 34 and 47~km/s/kpc \citep[][and references therein]{BlandHawthorn16}, the ILR is expected to be located out of our region of study.
%
%
%
%
%
%
\section{Conclusions}
\label{Sec_conclusion}
\par The statistics of the radial actions reveal arc-shape structures in the Galactic disc. These structures are characterised by a predominance of more circular orbits that contrasts to the high radial action feature found at $R\sim10.5$~kpc.
\par The analysis of the errors in $J_R$ confirms the reported structures are not spurious but robust from the statistical point of view. Furthermore, they cannot be explained by the selection effects inherent in \textit{Gaia}.
\par The characteristic arc shape of the structures in $J_R$ motivates the comparison with the Milky Way spiral arms, whose fit parameters have been reported in previous studies. We find that, in the innermost region, Structure A clearly defines the Sagittarius arm, with its upper boundary is delimited by the Scutum arm. At larger Galactocentric radii, Structure B tracks the Local Arm at negative $X$ while no clear correspondence with literature is found at $X>0$, where the variety of models suggests a complex definition for this arm. On the contrary, for the Perseus Arm we observe a good concordance with the spatial distribution of young stellar population for $X\in(-2, 0)$~kpc, while at positive $X$ the orientation of the $J_R$ feature has a different pitch angle compared to all the considered models. Our results suggest that the Perseus Arm in the $J_R$ map is connected to the Local Arm at $\sim 3.6$~kpc from the Sun, in the direction $\ell\approx-100^\circ$. This would result in a mismatch with some geometries of the spiral arms from young stellar populations, which will be studied in the future. We observe a correspondence between the segment 18 in \citet{Lemasle22} and a region of very low $J_R$ between Structures A and B that has not clear spiral arm assignation.
\par We also explore the moving groups as a possible explanation for the features. The $J_R$ arc-shape structures in the $(X,Y)$ plane are likely related to the structures in $J_R$ in the $R$-$V_{\phi}$ plane but mapped into different projections of phase space, in particular showing also their complex dependency with position (e.g. azimuth) in the $(X,Y)$ case. We observe some features in the $V_{\phi}$ vs. $R$ plane that might be anti-correlated with some known moving groups. However, this connection between the moving groups and the $J_R$ features in the Galactic plane, if present, is not obvious and should be explored in future studies. 
\par We identify an area of high radial action centered at $\sim$~$10.5$~kpc, where the Outer Lindblad Resonance (OLR) caused by the bar is expected. Apart from the features in the maps of the radial action, we find the distribution of the giant stars in the disc is consistent with the spiral arms traced by younger populations; in particular, the upper main sequence stars.
\par The analysis presented in this work indicate that multiple agents might be causing the structures found in the distribution of $J_R$. Although the spiral arms account for most of the features reported in this work, there are still many discrepancies that must be addressed. In this context, further studies with numerical simulations and analytical models are required to explain these differences and shed light on the Galactic dynamics.
\begin{acknowledgements}
The authors acknowledge J. Bland-Hawthorn for his constructive contribution to this work as referee. We thank P. de Laverny for his useful comments. P. A. Palicio acknowledges the financial support from the Centre national d'études spatiales (CNES). E. Spitoni and A. Recio-Blanco received funding from the European Union’s Horizon 2020 research and innovation program under SPACE-H2020 grant agreement number 101004214 (EXPLORE project). This project has received funding from the European Union's Horizon 2020 research and innovation programme under the Marie Sklodowska-Curie grant agreement N. 101063193. TA acknowledges the grant RYC2018-025968-I funded by MCIN/AEI/10.13039/501100011033 and by ``ESF Investing in your future''. This work was (partially) funded by the Spanish MICIN/AEI/10.13039/501100011033 and by ``ERDF A way of making Europe'' by the ``European Union'' through grant RTI2018-095076-B-C21, and the Institute of Cosmos Sciences University of Barcelona (ICCUB, Unidad de Excelencia 'Mar\'{\i}a de Maeztu') through grant CEX2019-000918-M. PJM acknowledges project grants from the Swedish Research Council (Vetenskapr{\aa}det, Reg: 2017- 03721; 2021-04153). This work has made use of data from the European Space Agency (ESA) mission {\it Gaia} (\url{https://www.cosmos.esa.int/gaia}), processed by the {\it Gaia} Data Processing and Analysis Consortium (DPAC, \url{https://www.cosmos.esa.int/web/gaia/dpac/consortium}). Funding for the DPAC has been provided by national institutions, in particular the institutions participating in the {\it Gaia} Multilateral Agreement. Although \textsc{galpy} is not explicitly used in this work, P. A. Palicio uses its source code as reference and recognises the credit for the work of \citet{Bovy14}.
\end{acknowledgements}


\bibliographystyle{aa}  
\bibliography{biblio} 
\begin{appendix}
\section{\texttt{ADQL} query}
\label{Appendix:Query}
\lstset{language=SQL}
\begin{lstlisting}[caption={\texttt{ADQL} query for the \textit{Gaia} DR3 considered in this work.},captionpos=b]
SELECT source_id, ra, dec, pmra, pmdec, radial_velocity, parallax, ruwe, ra_error, dec_error, pmra_error, pmdec_error, radial_velocity_error, parallax_error, ra_dec_corr, ra_pmra_corr, ra_pmdec_corr, dec_pmra_corr, dec_pmdec_corr, pmra_pmdec_corr, grvs_mag,r_med_geo, r_lo_geo, r_hi_geo
FROM user_dr3int6.gaia_source INNER JOIN external.gaiaedr3_distance USING(source_id)
WHERE (radial_velocity is not NULL) and (pmra is not NULL) and (pmdec is not NULL) 
\end{lstlisting}

\section{Stäckel-Fudge approximation}
\label{AppendixStackel}
Within this approach, the orbital parameters can be computed assuming the considered Galactic potential $\Phi(R,z)$ satisfies some properties of the so-called Stäckel potentials. Given an axisymmetric oblate distribution of mass, its potential $\Phi(R,z)$ is said to be a Stäckel potential if there are two single-variable functions $U(u)$ and $V(v)$ such that
\begin{equation}
\label{Eq_UV}
\Phi_S(u,v)=\frac{U(u)-V(v)}{\sinh^2{u}+\sin^2{v}}
\end{equation}
where $(u, v)$ are the ellipsoidal coordinates \citep{deZeeuw1985} related to $(R, z)$ through the transformation
\begin{equation}
\label{Eq_transform}
R = \Delta \sinh{u}\sin{v}\ \ \ \ \ \ \ \ \ \ z = \Delta \cosh{u}\cos{v}
\end{equation}
with $\Delta$ the focal length of the elliptical (hyperbolic) curves of constant $u$ ($v$). Since the Galactic potential is known to be oblate, we do not describe the prolate case \citep[for the prolate case see ][]{deZeeuw1985}. By differentiating both sides of Eq. \ref{Eq_transform} with respect to time, the transformation of the momentum between (R, z) and the (u, v) coordinate system results
\begin{eqnarray} \nonumber
p_u &=&p_R\Delta\cosh{u}\sin{v} + p_z\Delta\sinh{u}\cos{v} \\
p_v &=& p_R\Delta\sinh{u}\cos{v}- p_z\Delta\cosh{u}\sin{v}
\label{Eq_pupv}
\end{eqnarray}
where $p_i$ is the momentum associated with the coordinate $i\in \lbrace{ R, z, u, v\rbrace}$. The Hamiltonian constructed with the momenta of Eq.\ref{Eq_pupv} and the potential $\Phi_S(u,v)$ results in a expression that can be separated into two single variable terms:
\begin{eqnarray}
\label{Eq_Esh}\nonumber
E\sinh^2{u} &=& \frac{p_u^2}{2\Delta^2}+U(u)+I_3+\frac{L_z^2}{2\Delta^2 \sinh^2{u}}\\
E\sin^2{v} &=& \frac{p_v^2}{2\Delta^2}-V(v)-I_3+\frac{L_z^2}{2\Delta^2 \sin^2{v}}
\end{eqnarray}
in which $E$ is the total energy of the system (since the Hamiltonian does not depend explicitly on time), $L_z$ is the vertical component of the angular momentum and $I_3$ is the third integral of motion.

\par For a reference point with coordinates $(u, v)=(u_0, \pi/2)$ the expression for $u$ in Eq. \ref{Eq_Esh} reads
\begin{eqnarray}
\label{Eq_Esh0}
E\sinh^2{u_0} &=& \frac{{p_u^2}_0}{2\Delta^2}+U(u_0)+I_3+\frac{{L_z}^2}{2\Delta^2 \sinh^2{u_0}}\\
E &=& \frac{p_{0.5\pi}^2}{2\Delta^2}-V(\pi/2)-I_3+\frac{{L_z}^2}{2\Delta^2}
\end{eqnarray}
where the choice for $u_0$ is discussed later. Subtracting Eq. \ref{Eq_Esh} from \ref{Eq_Esh0} and solving for $p_u$ we find
\begin{eqnarray}
\nonumber
\label{Eq_for_pu}
\frac{p_u^2}{2\Delta^2} = \frac{{p_u}^2_0}{2\Delta^2} + E\left(\sinh^2{u}-\sinh^2{u_0}\right) -U(u)+U(u_0)\\
-\frac{{L_z}^2}{2\Delta^2} \left(\frac{1}{\sinh^2{u}}-\frac{1}{\sinh^2{u_0}}\right)
\end{eqnarray}
where the term $\delta U \equiv U(u)-U(u_0)$ can be approximated using the definition of the Stäckel potential (Eq. \ref{Eq_UV}) as
\begin{eqnarray}\nonumber
\label{Eq_for_deltaU}
\delta U \equiv U(u)-U(u_0) \approx (\sinh^2{u} +\sin^2{v})\Phi(u,v)\\-\left(\sinh^2{u_0} + \sin^2{v}\right)\Phi(u_0,v)
\end{eqnarray}
Similarly, we can define $\delta V = V(v)-V(\pi/2)$ such that
\begin{eqnarray} \nonumber
\label{Eq_for_deltaV}
\delta V &\equiv& \cosh^2{u}\Phi(u, \pi/2)-\\
&-&(\sinh^2{u} + \sin^2{v})\Phi(u, v),
\end{eqnarray}
then $p_v$ can be written as
\begin{equation}
\label{Eq_for_pv}
\frac{p_v^2}{2\Delta^2} = E\sin^2(v) + I_3+V(\pi/2) + \delta V - \frac{L_z^2}{2\Delta^2 \sin^2{v}}.
\end{equation}
The expressions \ref{Eq_for_pu} and \ref{Eq_for_pv} for $p_u$ and $p_v$ respectively can be substituted in the integrals for the definition of the actions \citep[see Eq. 6 in ][]{Binney12}
\begin{eqnarray}
\label{Eq_Ju_Jv}
J_u &=& \frac{1}{2\pi} \oint p_u du =  \frac{1}{\pi} \int_{u_{min}}^{u_{max}} p_u du\\
J_v &=& \frac{1}{2\pi} \oint p_v dv = \frac{2}{\pi} \int_{v_{min}}^{\pi/2} p_v dv
\end{eqnarray}
\par
\begin{figure}[ht]
\centering
\includegraphics[width=0.32\textwidth]{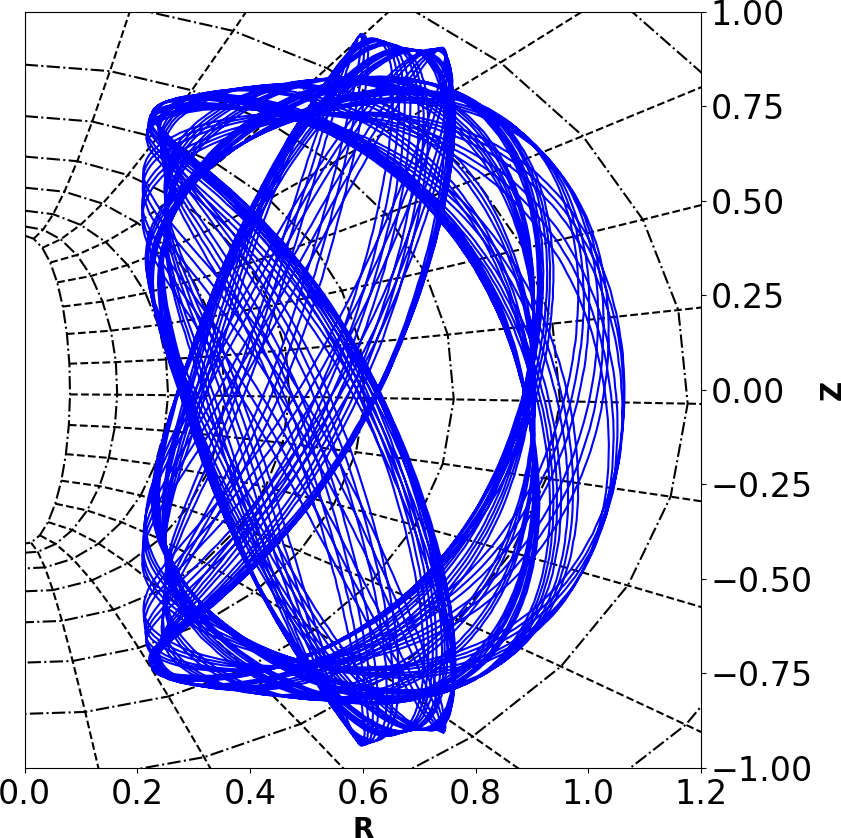}
\caption{Example of an orbit in the ($R$, $Z$) plane (blue curve) with the lines of constant $u$ (dot-dashed ellipses) and $v$ (dashed hyperbolas) in the background. The units of the axis are arbitrary.}
\label{Fig_uvexample}
\end{figure}

 Both the limits and the integrals of Eq. \ref{Eq_Ju_Jv} have to be computed numerically. The limits of integration correspond to the roots of Eq. \ref{Eq_for_pu} and \ref{Eq_for_pv} (therefore the actions are always real). In our case, we compute these roots using the bisection method while the numerical integration is performed by Gaussian Quadrature with ten nodes. We approximate the radial and vertical actions as $J_R\approx J_u$ and $J_z\approx J_v$, respectively, since the $R$ ($z$) coordinate varies more with $u$ ($v$), as Fig \ref{Fig_uvexample} illustrates. The choice of $u_0$ is rather arbitrary (see Section 2 in \citet[][]{Binney12} for the discussion), so we use the coordinate $u$ given by the input value $(R,z)$ of the star.
%
%
\par In order to account for the error propagation, we perform 25 random realisations of the input data and compute the median values and the 16-th and 84-th percentiles of the output.
\section{Results with the subsample of giants}
\label{Appendix:Giants}
\par
\begin{figure*}[]
\includegraphics[width=0.99\textwidth]{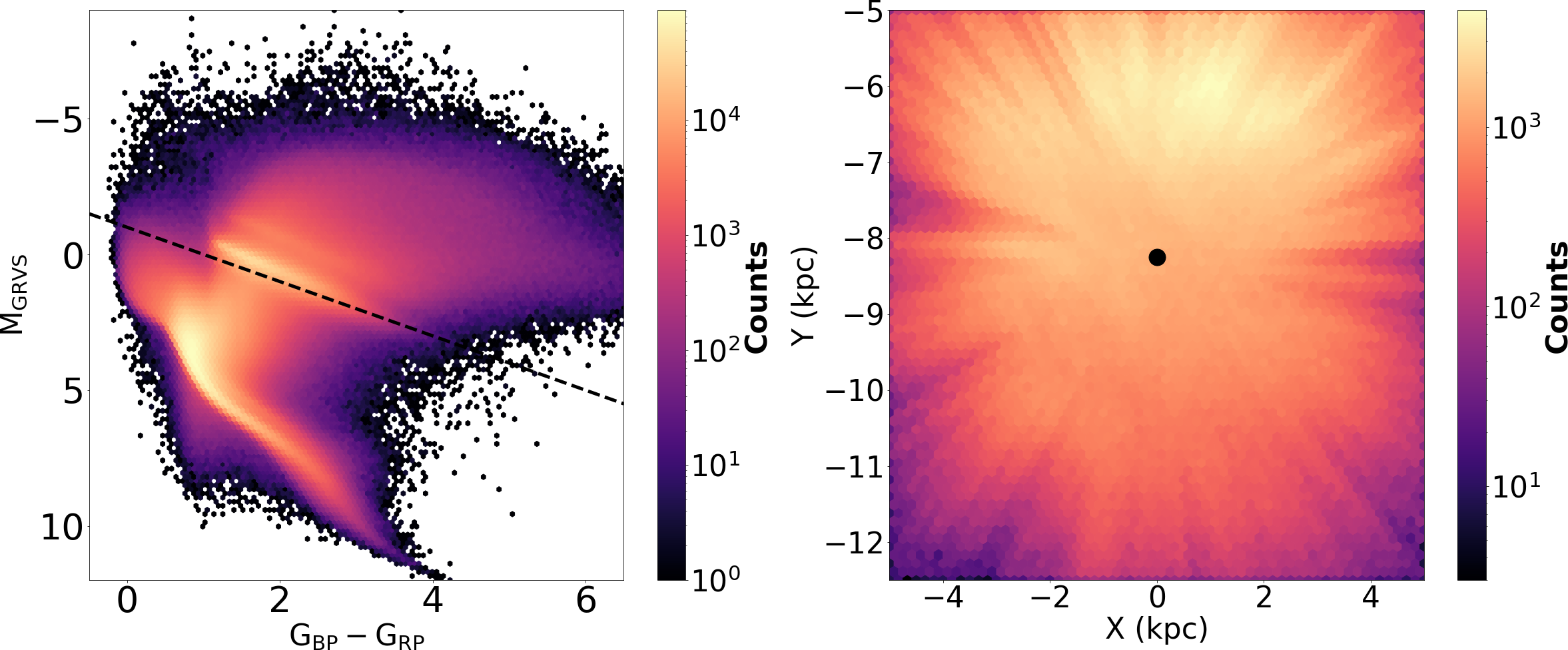}
\caption{Left panel: distribution of stars in the Hertzsprung–Russell diagram for the Galactic disc sample ($|Z_{max}|<0.5$~kpc). The dashed black line represents the boundary condition considered to separate Main Sequence and RGB stars (Eq. \ref{Eq_photcriteria}) neglecting extinction. Right panel: density map of the giant sample in the $(X, Y)$ plane. The solar position is denoted by a solid black circle.}
\label{HR_giant_selection}
\end{figure*}

 Using a photometric selection of stars, we demonstrate that the dynamical pattern reported here is mainly supported by the old giant population, although the stars in spiral arms tend to be younger than average and, therefore, have lower values of $J_R$ according to the age-velocity dispersion relation. We reproduce the percentiles of $J_R$ shown in Figure \ref{Fig_percentJR} applying the following photometric criteria
\begin{equation}
    \label{Eq_photcriteria}
    G_{RVS}+5-5\log_{10}d_{pc} < G_{BP}-G_{RP} - 1
\end{equation}
which assumes no extinction as a first approximation to restrict the sample to the giants (hereafter, we refer to this subset as \textit{giant subsample}). The expression in Eq. \ref{Eq_photcriteria} visually separates the Red Giant Branch (RGB) from the Main Sequence stars in the Hertzsprung–Russell (HR) diagram using the Red Clump as reference for the boundary (Fig. \ref{HR_giant_selection}). By selecting giants we keep stars intrinsically brighter, and reduce the effect of the selection function and the contribution of the faint dwarf stars that dominate the sample in the Solar neighbourhood \citep{PVPCCoMW}.
\par
\begin{figure*}[]
\includegraphics[width=0.98\textwidth]{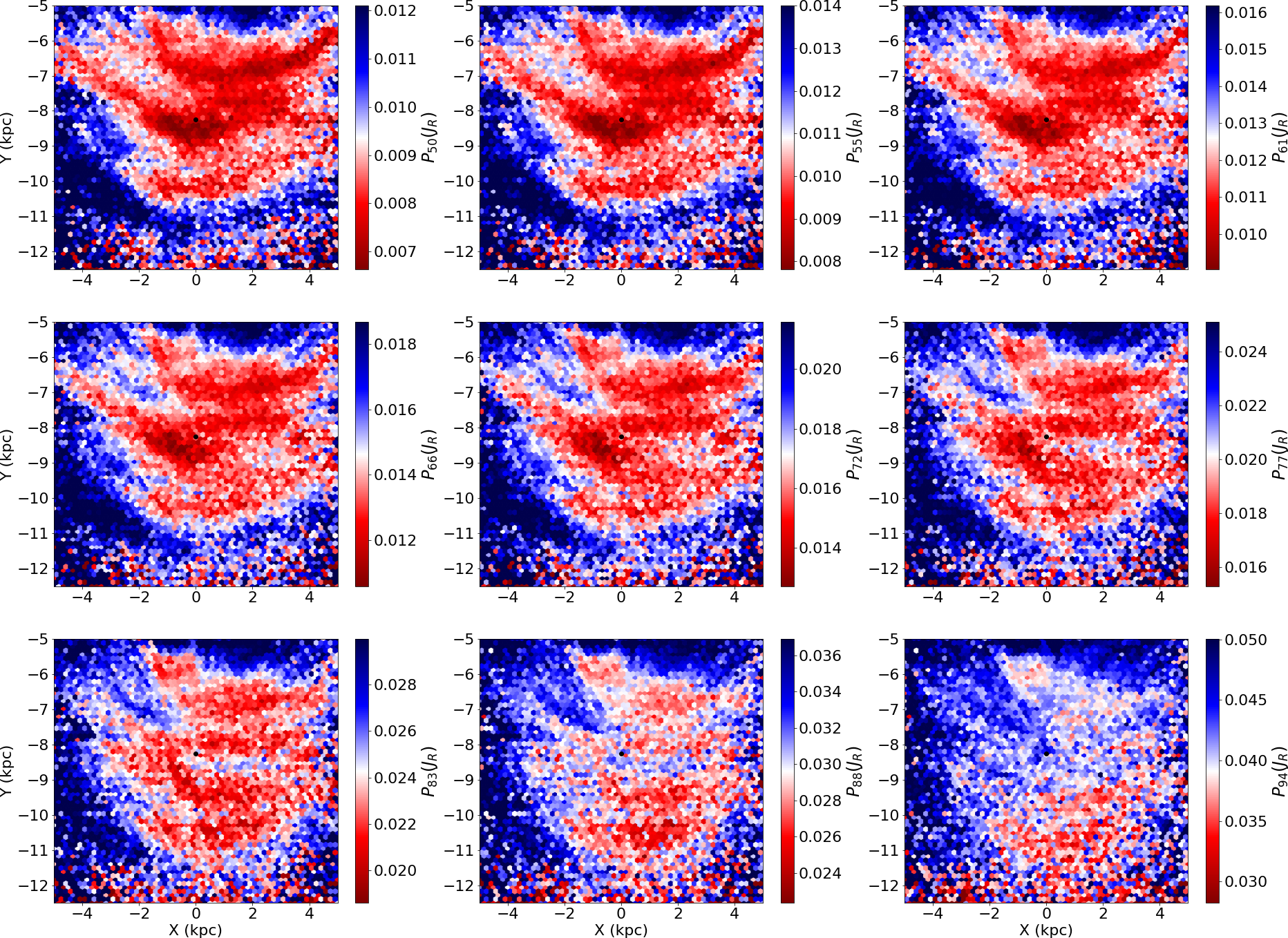}
\caption{Similar to Figure \ref{Fig_percentJR} but for the \textit{giant subsample}. The upper left panel corresponds to the median (equivalent to Figure \ref{Fig_medianJR}) while two additional percentiles are shown in the central and right upper panels.}
\label{Fig_percentJR_giants}
\end{figure*}

 Figure \ref{Fig_percentJR_giants} illustrates the distribution of the percentiles of $J_R$ for the \textit{giant subsample}. In general, the features found in the whole sample are observed in the \textit{giant subsample}, with the exception of the high $J_R$ region between the Local and Perseus (B-C) arm which is more distorted. Similarly, the high radial action region near the Sun disappears. In contrast to Figure \ref{Fig_densities}, for the \textit{giant subsample} the highest density area corresponds to the innermost low $J_R$ region, though no evidence of the other structures are observed.
%
%
\par
\begin{figure*}[]
\includegraphics[width=0.98\textwidth]{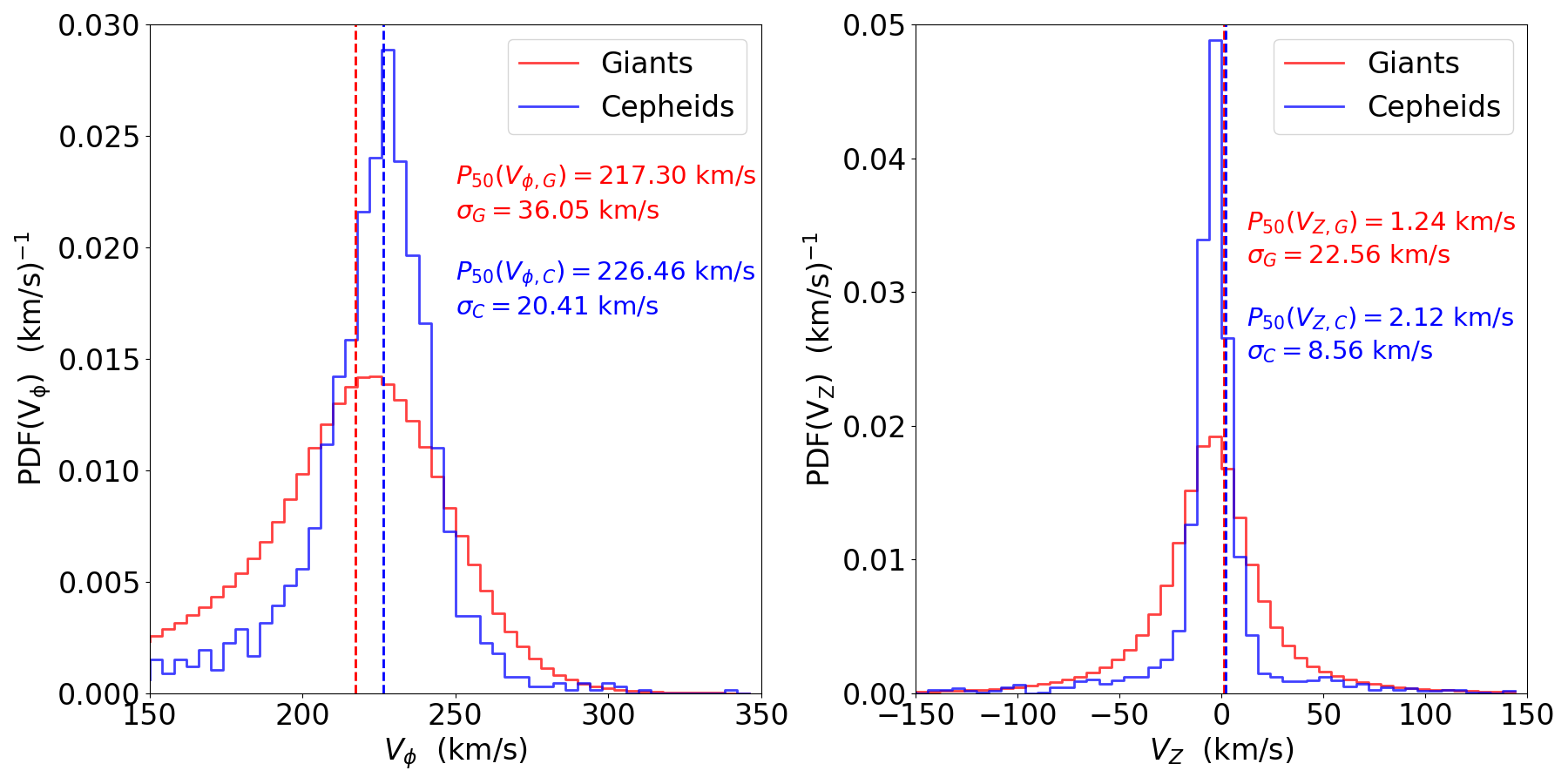}
\caption{Probability distribution function of the azimuthal (left panel) and vertical (right panel) velocities for the subsample of giant stars (red bars) and Cepheids (blue bars). The median values ($P_{50}$) and median absolute deviations ($\sigma$) for both samples are indicated in the inset. Vertical dashed lines denote median values.}
\label{Fig_histogram}
\end{figure*}

 We perform an additional test to address the age of the tracers of the arc-shaped structures. We compare the kinematics of our \textit{giant sample} with that of the Cepheids in \textit{Gaia} DR3 \citep{Ripepi22} to get a proxy of the relative age of both populations. In Figure \ref{Fig_histogram}, we show the distributions of azimuthal ($V_{\phi}$) and vertical velocities ($V_Z$) for the Cepheid and \textit{giant} subsamples. As we can see, the distribution of $V_{\phi}$ for the Cepheids is more peaked than that of the \textit{giants}, as expected for cooler (and younger) populations, with a median absolute deviation\footnote{Defined as $\sigma=1.48\times\textnormal{median}(|x-\textnormal{median}(x)|)$ for consistency with the standard deviation.} of $\sigma_C=36.05$~km/s ($\sigma_G=20.41$~km/s) for the Cepheid (giant) samples. Similarly, the distribution of $V_Z$ observed for the \textit{giant subsample} is consistent with a hotter and older population compared to the Cepheids, with $\sigma_G=22.56$~km/s and $\sigma_C=8.56$~km/s, respectively. Therefore, the distributions presented in Fig. \ref{Fig_histogram} indicate that the \textit{giant subsample} is dominated by stars typically older than the Cepheids.
\par It is interesting to note that our sample is typically old, while the spiral arms are often seen in young stars (classical Cepheids, masers, UMS stars). When comparing the features in $J_R$ to the spiral arms, therefore, we should bear in mind that intrinsic differences might be present because they are two different populations.
%
%
%
\end{appendix}
\end{document}